\documentstyle[aps,prb,floats,twocolumn,epsf]{revtex}

\begin{document}

\twocolumn[\hsize\textwidth\columnwidth\hsize\csname
@twocolumnfalse\endcsname

\title{Quantum Theory of Quantum-Hall Smectics}

\draft

\author{A.H. MacDonald$^{1,2}$ and Matthew P.A. Fisher$^{2}$}
\address{$^{1}$Department of Physics, Indiana University, Bloomington, IN 
47405-4202}
\address{$^{2}$ Institute for Theoretical Physics, University of California at
Santa Barbara, Santa Barbara CA 93106-4030}
\date{\today}
\maketitle

\begin{abstract}
We propose a quantum stripe (smectic) coupled-Luttinger-liquid model for the
anisotropic states which occur in two-dimensional electron systems with
high-index partial Landau level filling, $\nu^{*} = \nu - \lbrack\nu\rbrack$.
Perturbative renormalization group calculations establish that interaction terms
neglected in this model are relevant - probably driving the system into an
anisotropic Wigner crystal---but for $0.4 \lesssim \nu^{*} \lesssim 0.6$ only
below temperatures which are outside of the experimentally accessible range. We
argue that the Hall conductance of the ground state flows toward
$\lbrack\nu\rbrack e^{2}/h$ and $(\lbrack\nu\rbrack + 1) e^{2}/h$ respectively, on
the low and high filling factor sides of this range, consistent with recent
observations. A semiclassical theory of smectic state transport properties,
which incorporates Luttinger liquid effects in the evaluation of scattering
amplitudes, accounts for the magnitude of the dissipative resistivities at
$\nu^{*}=1/2$, for their $\nu^{*}$-dependence, and for the observation of
non-linearities of opposite sign in easy and hard direction resistivities. 
\end{abstract}
\vskip2pc]

\narrowtext 
 
\section{Introduction}

Recent transport experiments\cite{lilly,du,shayegan} have established a
qualitative difference between low-energy states of two-dimensional electron
systems with large and small index partially filled Landau levels. For Landau
level filling factors $\nu < 4$ (orbital Landau level indices smaller than
$N=2$), isotropic quantum Hall fluid states occur at fractional values of $\nu$.
For $ N \ge 2$, on the other hand, experiments have discovered regions of
strongly anisotropic dissipative transport near half-odd-integer filling
factors, bracketed by reentrant integer-quantum-Hall effect regions with Hall
conductivities $\lbrack\nu\rbrack (e^{2}/h)$ and $(\lbrack\nu\rbrack
+1)(e^{2}/h)$. This dependence on $N$ is presumably due to subtle changes in the
effective interactions among the electrons of the partially filled Landau level.
In this article we describe a theory which accounts qualitatively and often
semi-quantitatively for the principle facts uncovered by this series of
experiments. 

Following Eisenstein and co-workers,\cite{lilly} we start from the assumption
that the true ground state is close to the unidirectional charge-density-wave
states proposed for $N \ge 2$ on the basis of Hartree-Fock calculations by
Koulakov {\em et al.}\cite{fogler} and Moessner and Chalker.\cite{moessner} In
Section II we derive a model of coupled one-dimensional chiral Luttinger liquid
electron systems for this state. The derivation provides microscopic expressions
for the interaction parameters of the model, which are long-range because of the
long-range of the underlying Coulomb interaction between electrons. This model
neglects small interstripe backscattering terms. In Section III we demonstrate
that these terms are technically relevant, but near half filling ($0.4 \lesssim
\nu^{*} \lesssim 0.6$) only at inaccessibly low temperatures. Outside this
range, however, observable Wigner crystal instabilities are predicted. In
Section III we present an estimate of the $\nu^{*}$ dependence of the
temperature below which Wigner crystal states are expected to form. Transport
physics in the interesting stripe state regime near $\nu^{*} = 1/2$ is considered
in Section IV. We present a semiclassical theory in which Luttinger liquid
effects are incorporated in the evaluation of scattering amplitudes and which
describes experiments\cite{lilly,du,shayegan} semi-quantitatively. This theory
makes a number of parameter free quantitative predictions which are in good
accord with observations. In particular, the product of easy and hard direction
resistivities in this theory is independent of disorder strength and has a value
which agrees well with experiments. Moreover, Luttinger liquid effects lead to a
natural explanation of the non-linear transport effects observed experimentally.

Several recent papers\cite{rezayi,fertig,phillips,jungwirth} have explored the
properties of interacting electron systems in higher Landau levels. The basic
framework of our theory has much in common with the work of Fradkin and
Kivelson,\cite{fradkin} whose approach intriguingly suggests\cite{nature} a
similarity between the strong correlation physics of quantum Hall and doped Mott
insulator systems. These authors have emphasized the intimate relationship
(based on shared symmetry properties) between uni-directional charge density
wave states and smectic liquid crystal states. We have followed their lead in
referring to the anisotropic high Landau level states as quantum-Hall smectics.
Both theories identify the electron stripes as one dimensional electron systems
and use bosonization techniques to describe the low energy exctiations of their
left-going and right-going states. The most important difference in our work is
that stripe position and shape fluctuations are identified microscopically with
the same low energy excitations. They are {\em not} separate low-energy degrees
of freedom. Our theory can be developed in terms of either standard Luttinger
liquid boson fields or equivalently in terms of stripe width and position
fields. We find one set of gapless collective modes for quantum-Hall smectics,
which encompasses all of the low energy degrees of freedom. A physical
consequence of this difference is that in our theory, the quantum-Hall smectic
ground state is {\em always unstable} to the formation of either an electron or
a hole Wigner crystal, depending on the sign of $1/2-\nu^{*}$. 

\section{Quantum Smectic Model}

\begin{figure}
\begin{center}
\epsfxsize=8.5cm
\epsffile{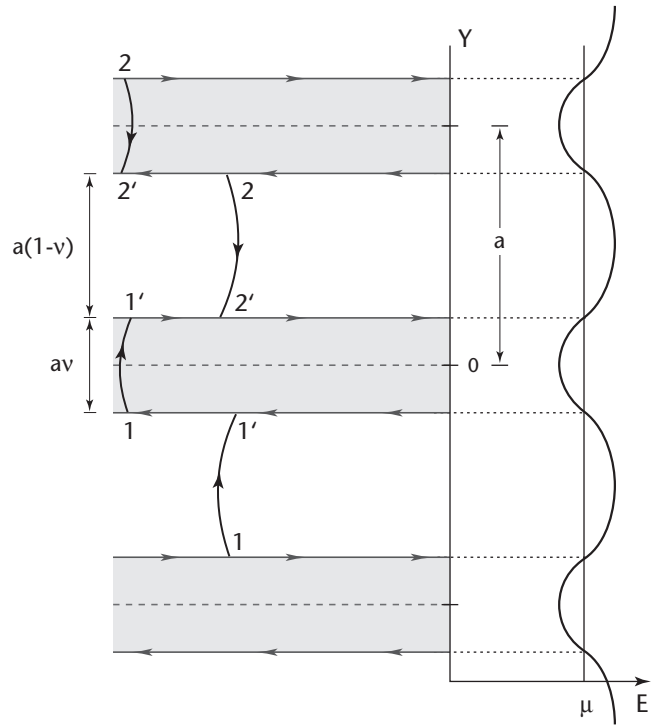}
\end{center}
\caption{Schematic illustration of the Hartree-Fock theory smectic state. This
state is a local minimum of the Hartree-Fock energy functional for any value of
$\nu^{*}$ and any Landau level index. At filling factor $\nu^{*}$, the occupied
Landau gauge single-particle states have guiding centers in stripes of width $a
\nu^{*}$, shaded in this figure, which repeat with period $a$. The state can be
viewed as consisting of periodically repeated electron stripes or hole stripes.
The Hartree-Fock single-particle eigenvalues lie below the Fermi level for
guiding centers in the stripes and above the Fermi level for guiding centers
outside the stripes. We take the $\hat x$ direction to be along the stripes and
the $\hat y$ direction to be across the stripes. In a magnetic field, the
guiding center is related to wavevector by $k = y / \ell^{2}$ where $\ell$ is
the magnetic length. Each stripe has right and left-going Hartree-Fock
quasiparticles at its top and bottom edges respectively. In the Luttinger liquid
theory for the one-dimensional stripes, the local Fermi momentum for left and
right going states in each stripe is elevated to a quantum field. Because of the
connection between guiding center and momentum, these fields also describe the
thermal and quantum fluctuations of the shapes and positions of the electron and
hole stripes. The number of right and left going states in any channel is
related to its Fermi wavevector by $\rho_{n,\pm} = k_{F,\pm}/ 2 \pi$. The
strongest momentum-conserving interaction terms not included in the
non-interacting boson limit of the Luttinger liquid theory are those in which
electrons scatter from left to right-going states in one electron stripe or hole
stripe and from right to left-going states in a different stripe of the same
type.}
\label{fig:one}
\end{figure}

The smectic state of Hartree-Fock theory\cite{fogler,moessner} is a
single-Slater-determinant with alternating occupied and empty guiding-center
occupation-numbers stripes as illustrated schematically in Fig.~(\ref{fig:one}).
These states spontaneously break translational and rotational symmetry. For
large $N$ they tend to have lower energy than isotropic fluid states because the
electrostatic energy penalty, which usually thwarts the
phase-separation\cite{nature} favored by exchange interactions and by electronic
correlations, is small\cite{fogler,moessner} when the density wave period is
comparable to the cyclotron orbit diameters of index $N$ electrons. We can
consider these states to be composed of either electron or hole stripes with
right and left going quasiparticles at opposite edges.

Small fluctuations in the positions and shapes of the stripes can be described
in terms of particle-hole excitations near the stripe edges. The residual
interactions, ignored in Hartree-Fock theory, which scatter into these low
energy states fall into two classes: ``forward" scattering interactions which
conserve the number of electrons on each edge of every stripe, and ``backward"
scattering processes which do not. The latter processes involve large momentum
transfer and will be smaller in magnitude (see below). The quantum smectic model
described in this section includes forward scattering only. These interactions
are bilinear in the 1d electron densities associated with the chiral currents at
the stripe edges: $\rho_{n\alpha}(x)$, with $\alpha=\pm$. As explained in
Fig.~(\ref{fig:one}), these densities are proportional to an ``elastic" field
$u_{n\alpha}(x) = \alpha 2 \pi \ell^{2} \rho_{n\alpha}(x)$ (with $\ell = (\hbar
c/eB)^{1/2}$ the magnetic length), which measures the transverse displacement of
a stripe edge relative to its presumed equilibrium position, $y_{n\pm}^{0} = a(n
\pm \nu^{*}/2)$. The quadratic Hamiltonian which describes the {\em classical}
energetics for small fluctuations has the following general form:
\begin{eqnarray}
H_{0} &=& \frac{1}{2\ell^{2}} \int_{x,x'} \sum_{n,n'} u_{n\alpha}(x) D_{\alpha
\beta}(x-x';n-n') u_{n'\beta}(x') \nonumber \\
&=& \frac{1}{2\ell^{2}} \int_{\bbox{q}} u_{\alpha}(- \bbox{q}) D_{\alpha
\beta}(\bbox{q}) u_{\beta}(\bbox{q}) ,
\label{hamiltonian}
\end{eqnarray}
where $\int_{\bbox{q}} \equiv \int d^{2}{\bbox{q}}/(2 \pi)^{2}$. Here the
$q_{y}$ integral is over the interval $(-\pi/a,\pi/a)$ and a high momentum
cutoff $\Lambda \sim 1/\ell$ is implicit on $q_{x}$. 

Symmetry considerations further constrain the form of the elastic kernel. In
position space the kernel must be real and symmetric so that, $D_{\alpha
\beta}(\bbox{q}) = D^{*}_{\alpha \beta} (-\bbox{q}) = D^{*}_{\beta
\alpha}(\bbox{q})$. This implies $D_{-+}(\bbox{q})=D^{*}_{+-}(\bbox{q})$ and
${\rm Im} D_{\alpha \alpha}(\bbox{q}) = 0$. Parity invariance (under $x,n,+
\leftrightarrow -x,-n,-$), implies moreover $D_{++}(\bbox{q})=D_{--}(\bbox{q})$.
Thus, the elastic kernel is fully specified by one real function,
$D_{++}(\bbox{q})$, and one complex function, $D_{+-}(\bbox{q})$. In the {\em
g-ology} notation of the 1D electron gas literature\cite{1drefs,soloym} these
amplitudes correspond to $g_{4}$ and $g_{2}$, respectively. Finally, provided
the broken translational and rotational invariance in the smectic occur
spontaneously, the classical Hamiltonian must be invariant under:
$u_{n\alpha}(x) \rightarrow u_{n\alpha}(x)+ {\rm const}$ and $\partial_{x}
u_{n\alpha}(x) \rightarrow \partial_{x} u_{n\alpha}(x)+ {\rm const}$. This
symmetry determines the form of $D(\bbox{q}) = \sum_{\alpha \beta} D_{\alpha
\beta} (\bbox{q})$ at small wavevector:
\begin{equation}
D(\bbox{q}) = K_{x} q_{x}^{4} + K_{y} q_{y}^{2} + ....  ,
\end{equation}
the characteristic form for smectic elasticity.

A {\em quantum} theory of the quantum-Hall smectic is obtained by imposing
Kac-Moody commutation relations on the chiral densities:
\begin{equation}
[\rho_{n\alpha}(x),\rho_{n'\beta}(x')] = \frac{i}{2\pi} \alpha
\delta_{\alpha,\beta} \delta_{n,n'} \partial_{x} \delta(x-x').
\end{equation}
This commutator together with $H_{0}$ fully specifies the quantum dynamics.
Electron operators in the chiral edge modes are related to the 1d densities via
the usual bosonic phase fields: $\psi_{n\alpha} \sim e^{i\phi_{n\alpha}}$ with
$\rho_{n\alpha} = \alpha \partial_{x} \phi_{n\alpha} /2\pi$. It is a notable
feature of the strong field regime that the Luttinger liquid bosonic fields,
$\phi_{n,\pm}(x)$, fully determine the stripe position and shape fluctuations.
In terms of the bosonic fields, the local center of the $n^{th}$ stripe is 
\begin{eqnarray}
Y_{n}(x) &=&  a n + \frac{u_{n,+}(x) + u_{n,-}(x)}{2} \nonumber \\
&=& a n + \frac{\ell^{2} [\partial_{x} \phi_{n+}(x) 
+ \partial_{x} \phi_{n-}(x)]}{2} 
\label{position}
\end{eqnarray} 
and the local width of the $n^{th}$ stripe is 
\begin{eqnarray} 
W_{n}(x) &=& a \nu^{*} + u_{n,+}(x) - u_{n,-}(x) \nonumber \\
&=& a \nu^{*} + \ell^{2} [\partial_{x} \phi_{n+}(x) 
- \partial_{x} \phi_{n-}(x)] 
\label{width}
\end{eqnarray} 
We also remark that even though $H_{0}$ has a quadratic form, there is no limit
in which a free Fermion description of the smectic (with $D_{\alpha \beta}(n)
\sim \delta_{\alpha \beta} \delta_{n0}$) is valid. The interactions which are
responsible for the broken symmetry play an essential role.

Quantum properties of the smectic can be computed from the imaginary-time
action, 
\begin{eqnarray}
S_{0} &=& \int_{x,\tau} \frac{1}{4\pi} \sum_{n,\alpha} i \alpha \partial_{\tau}
\phi_{n,\alpha} \, \partial_{x} \phi_{n,\alpha} + \int_{\tau} H_{0} \nonumber \\
&=& \frac{1}{2} \int_{\bbox{q},\omega} \phi_{\alpha}(-\bbox{q},-\omega)
M_{\alpha,\beta} (\bbox{q},\omega) \phi_{\beta}(\bbox{q},\omega) ,
\label{action}
\end{eqnarray}
where in an obvious matrix notation,        
\begin{equation}
{\bf M}(\bbox{q},\omega) = (i \omega q_{x}/2\pi) \bbox{\sigma}^{z} + (q_{x}
\ell)^{2} {\bf D}(\bbox{q}).
\label{actionkernel}
\end{equation}
Correlation functions follow from Wick's theorem and the momentum space
correlator $\langle \phi_{\alpha} \phi_{\beta} \rangle = \bbox{M}^{-1}$ with
\begin{equation}
\bbox{M}^{-1}(\bbox{q},\omega) = \bbox{\sigma}_{z} \bbox{M}(\bbox{q},-\omega)
\bbox{\sigma}_{z} / {\rm det} \bbox{M}(\bbox{q},\omega) .
\label{eq:eight}
\end{equation}

Due to the spontaneous breaking of translational and rotational symmetry in the
smectic, one expects gapless Goldstone modes at zero wavevector. The collective
mode dispersion is readily obtained by setting ${\rm det}
\bbox{M}(\bbox{q},i\omega_{\bbox{q}})=0$ giving $\omega_{\bbox{q}} = v(\bbox{q})
q_{x}$ with a velocity,
\begin{equation}
v(\bbox{q}) = (2\pi \ell^{2}) [D_{++}^{2}(\bbox{q}) -
|D_{+-}(\bbox{q})|^{2}]^{1/2} .
\end{equation}
At small wavevectors, the mode velocity vanishes: $v^{2}(\bbox{q}) \sim
q_{y}^{2} + q_{x}^{4}$. Internal consistency requires that these soft modes do
{\em not} restore the symmetries assumed to have been broken in the smectic
state. To examine this we consider the complex smectic order parameter, $\Phi
\sim e^{iQu}$, which describes the charge-density order: $\delta \rho = Re \Phi
e^{iQy}$ with $Q=2\pi/a$ the ordering wavevector. The average $\langle \Phi
\rangle$ can be readily computed using the quantum harmonic theory, and at $T=0$
one finds $\langle \Phi \rangle \sim \exp(-Q^{2}I)$ with
\begin{equation}
I \sim \int_{\bbox{q}}  |q_{x}| D_{++}(\bbox{q}) /v(\bbox{q})   .
\end{equation}
With $D_{++}$ non-zero at $\bbox{q}=0$, the integrals converge at small
$\bbox{q}$, so that the deBye-Waller factor ($e^{- Q^{2} I}$) and smectic order
parameter are non-vanishing. Evidently, these harmonic quantum fluctuations are
insufficient to destroy the broken symmetries in the smectic.\cite{Finite-T}

\begin{figure}
\begin{center}
\epsfxsize=8.5cm
\epsffile{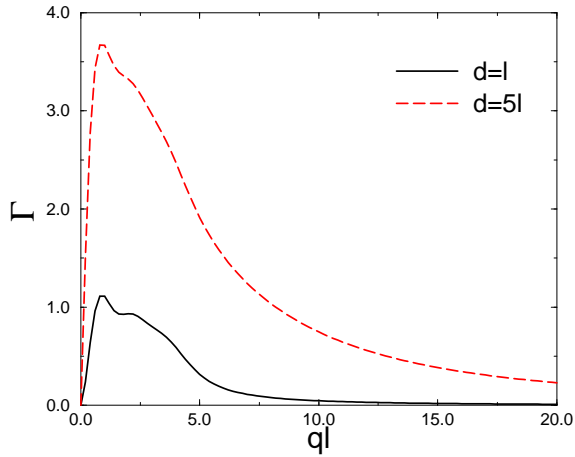}
\end{center}
\caption{Interaction matrix element $\Gamma(y)$ vs. dimensionless separation
$y/\ell = q \ell$ for the case of interactions in a zero-width 2D layer screened
by a parallel metallic layer and with Landau level index $N=2$. $d$ is the
distance to the metallic layer and $\Gamma$ is in units of $e^{2}/\epsilon
\approx 200 \, \mbox{\rm meV nm}$ for 2D electron systems formed near the
surface of a GaAs crystal. The Luttinger model g-ology parameter which
characterizes interactions between stripes separated by $ n a$ in the 2D
electron layer is $\sim \Gamma( n a)$. $\Gamma(y)$ is simply related to the
elastic constants in terms of which the chiral Luttinger model is developed in
the text. Metallic screening layers are sometimes present in experimental
samples but are introduced here mainly as a convenience since $\Gamma$ varies
logarithmically with $y$ at large $y$ if they are not present and various sums
over stripe indices do not converge. The limit $d \to \infty$ can be taken at
the end of the calculation, if appropriate. $\Gamma$ vanishes for $y \to 0$
because its direct and exchange contributions cancel.}
\label{fig:two}
\end{figure}

The effect of the neglected backscattering interactions, considered in the next
Section, depends sensitively on the elastic constants {\em at} $q_{x}=0$. In
this limit the relevant excited states are simply Slater determinants with
straight stripe edges displaced from those of the Hartree-Fock theory ground
state. By evaluating the expectation value of the microscopic Hamiltonian in a
state with arbitrary stripe edge locations we find that
\begin{equation}
D_{\alpha \beta}(q_{x}=0,q_{y}) = \delta_{\alpha \beta} D_{0} +
\alpha \beta \frac{a}{4 \pi^{2} \ell^{2}} 
\sum_{n} e^{iq_{y}an} \Gamma(y^{0}_{n\alpha} - y^{0}_{0\beta}) ,
\label{elastic}
\end{equation}
where the value of the constant $D_{0}$ is such that $\sum_{\alpha\beta}
D_{\alpha\beta}(\bbox{q}=0) = 0$. Here, $\Gamma(y)$ is the interaction potential
between two electrons located in guiding center states a distance $y$ apart:
\begin{equation}
\Gamma(y) = U(0,y/\ell^{2}) - U(y/\ell^{2},0),
\label{gamma}
\end{equation}
\begin{equation}
U(q,k) = \int \frac{d p}{2 \pi} e^{- (q^{2}+p^{2})\ell^{2}/2} \, V_{\rm
eff}^{N}(q,p) e^{- i p k \ell^{2}}.
\label{matrixelement}
\end{equation}
The two terms in Eq.~(\ref{gamma}) are direct and exchange contributions. In
Eq.~(\ref{matrixelement}), $V_{\rm eff}^{N}(q,p)$ is the Fourier transform of
the effective 2D electron interaction which incorporates form-factors
\cite{jungwirth} dependent on the Landau level index $N$ and the ground
subband wavefunction of the host semiconductor heterojunction or quantum well.
The smectic states have relatively long periods proportional to the index $N$
cyclotron orbit radii. Explicit calculations\cite{fogler,jungwirth} show that $a
\gtrsim 6 \ell$ for $N \geq 2$. It follows that the exchange contribution to
$\Gamma(y)$ is small and that $\Gamma(y)$ decreases with stripe separation in
the relevant range. With unscreened Coulomb interactions, $\Gamma(y)$ diverges
logarithmically at large $y$, so it is convenient to introduce a metallic
screening plane. This changes the large $y$ behavior to $y^{-2}$, making the sum
over $n$ in Eq.~\ref{elastic} convergent. As shown below, however, we do not
find that our conclusions change qualitatively when a screening plane is absent.
In Fig.~(\ref{fig:two})
we plot $\Gamma(y)$ for $N=2$ for the cases of thin 2D electron systems
separated from metallic screening planes by $d = \ell$ and $d = 5 \ell$. Note
that $\Gamma(y)$ is monotonically decreasing with positive curvature in the
range of interest. 

\section{Back Scattering Interactions} 

We now consider the ``backward" scattering electron interactions, ignored above.
The bare matrix elements for these interactions will fall off exponentially with
increasing momentum transfer and with increasing separation between the
interacting stripes, so we choose here to focus on the smallest momentum
transfer. We discuss explicitly only the case of
backscattering\cite{remarkotherterm} across electron stripes and across hole
stripes, as illustrated schematically in Fig. 1. For a pair of stripes separated
by $m a$, backscattering across an electron stripe can be expressed in a
bosonized form,
\begin{equation}
S_{1} =  \int_{x,\tau} \sum_{n,m} u_{m} [ \exp\left(i
\theta_{n,m}(x,\tau)\right) + h.c. ] ,
\end{equation}
where 
\begin{equation}
\theta_{n,m} =  (\phi_{n,+}-\phi_{n,-}) - (\phi_{n+m,+} - \phi_{n+m,-}).
\label{elec2kf}
\end{equation}
Hole backscattering takes a similar form. Since the effects of backscattering
across electron and hole stripes are equivalent under a particle/hole
transformation ($\nu^{*} \leftrightarrow 1-\nu^{*}$) we focus exclusively on the
former.

The effects of backscattering can be deduced by implementing a simple
renormalization group (RG) scheme. Specifically, we integrate out ``fast" boson
modes $\phi$ in a shell, with $\Lambda/b < |q_{x}| < \Lambda$ and $\omega,
q_{y}$ unrestricted, and then rescale $q_{x}^{\prime} = b q_{x}$ and
$\omega^{\prime} = b \omega$ leaving $q_{y}$ unchanged. With an appropriate
rescaling of $\phi$, this RG transformation leaves the harmonic smectic action,
$S_{0}$, invariant. Stability of the smectic fixed point in the presence of
backscattering can be tested by considering the lowest order RG flow equation,
\begin{equation}
\partial u_{m} / \partial t = (2 - \Delta_{m}) u_{m} ,
\end{equation}
with $t = \ln{b}$. Using Eq.~(\ref{elec2kf}) and Eq.~(\ref{eq:eight}) we find
the following expression for the scaling dimension:
\begin{equation}
\Delta_{m}= 4 \int_{-\pi}^{\pi} \frac{d (qa)}{2 \pi} \sin^{2}(mqa/2) W(q_{x}=0,q) .  
\label{scale-dim}
\end{equation}
Here, $W$ is a ``weight" function,  
\begin{equation}
W(\bbox{q}) = \frac{[D_{++}(\bbox{q}) +
ReD_{+-}(\bbox{q})]}{[D_{++}^{2}(\bbox{q}) - |D_{+-}(\bbox{q})|^{2} ]^{1/2}} .
\end{equation}
If the scaling dimension $\Delta_{m} < 2$ the smectic phase is {\em unstable}.
Fortunately, $\Delta_{m}$ only depends on the elastic constants {\em at} $q_{x}
=0$, so that we can use the microscopic expressions discussed at the end of the
previous section for its evaluation.

\begin{figure}
\begin{center}
\epsfxsize=8.5cm
\epsffile{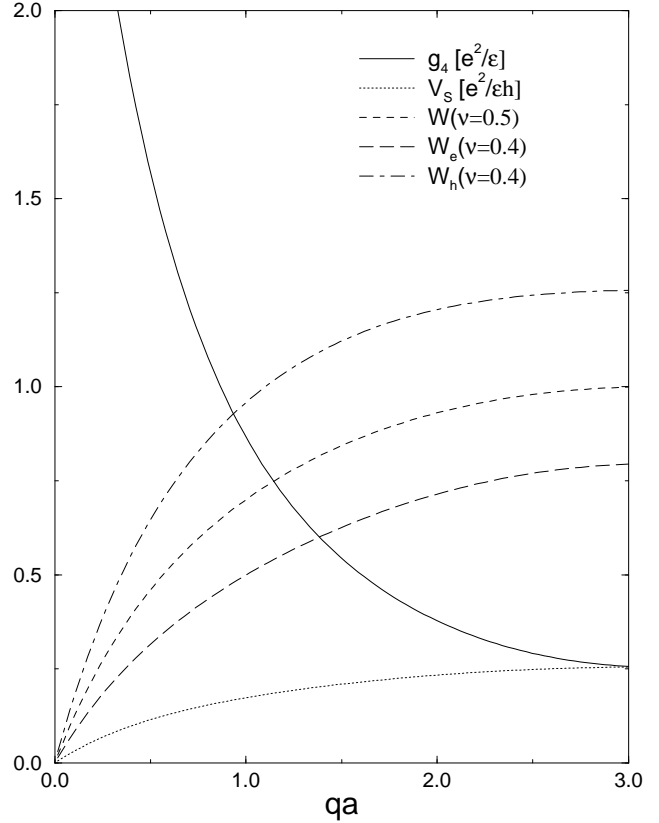}
\end{center}
\caption[]{Quantum-Hall smectic Luttinger model parameters and integrands
of the expressions for the backscattering interaction scaling dimensions. This plot is
for valence Landau level index $N=2$ and screening layer distance $d = 10 \ell$.
$g_{4}(q)$ is in units of $e^{2}/\epsilon$ and the collective excitation
velocity for $\nu^{*}=0.5$, $v_{S}(q_{x}=0,q_{y}q)$ is in units of $e^{2} / 2 \pi
\epsilon \hbar$. The values of these units are approximately $200 \mbox{\rm meV
nm}$ and $4.8 \times 10^{4} \mbox{\rm m/s}$ respectively for 2D electron systems
formed near the surface of a GaAs crystal. Our scaling dimension results can be
understood in terms of the properties of the weighting factors $W$ in the
integrals, as discussed in the text. $g_{4}(q)$ is related the elastic constants
in terms of which the chiral Luttinger model is developed in the text by
$D_{++}(k_x=0,q) = (a/4 \pi^{2} \ell^{2}) g_{4}(q)$. The large value of
$g_{4}(q)$ for $q \to 0$ is due to the long-range of the underlying Coulomb
interaction between electrons.}
\label{fig:three}
\end{figure}

\begin{figure}
\begin{center}
\epsfxsize=8.5cm
\epsffile{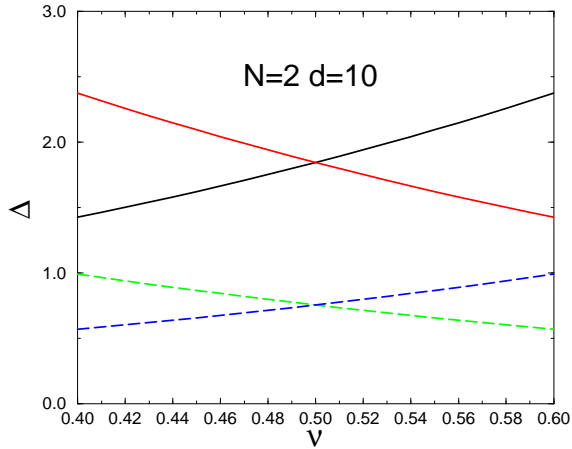}
\end{center}
\caption{Scaling dimensions for $m=1$ electron and hole $2k_{F}$ scattering and
electron and hole impurity scattering vertices (dashed lines) for a range of
filling factor near $\nu^{*} = 1/2$. For this calculation the distance to the
screening plane was chosen to be $d = 10 \ell$. Electron scattering vertices are
an increasing function of filling factor and hole vertices are a decreasing
function of filling factor as discussed in the text. The interaction terms are
relevant for scaling dimensions smaller than $2$ while impurity terms are
relevant for scaling dimensions smaller than $1.5$. The interedge scattering
rate is enhanced at low energies when the impurity interaction scaling dimension
is smaller than $1.0$. Interaction terms with $m$ larger than one are more
relevant but have bare coupling constants which are smaller by several orders of
magnitude. Interaction terms with larger momentum transfers than those discussed
here also have much smaller bare coupling constants.}
\label{fig:four}
\end{figure}

\begin{figure}
\begin{center}
\epsfxsize=8.5cm
\epsffile{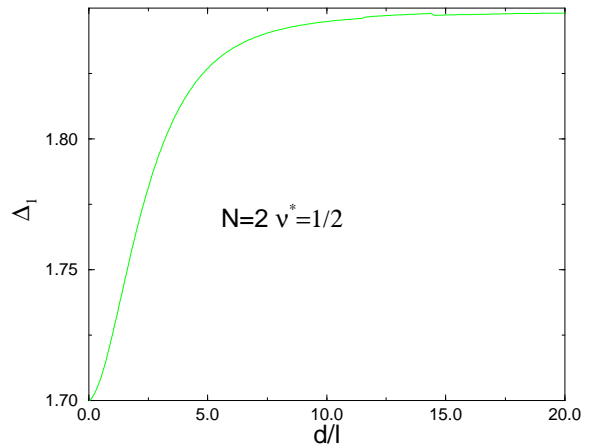}
\end{center}
\caption[]{Dependence of $m=1$ \, $\nu^{*}=1/2$ electron and hole backscattering
amplitude scaling dimensions ($\Delta_{1,e}=\Delta_{1,h}=\Delta_{1}$) on
distance between the two-dimensional electron system and the model's metallic
screening plane. Here $d$ is in units of the magnetic length $\ell$. For $d \to
0$, $\Delta$ approaches $16/3 \pi$, the value which can be calculated
analytically for the case of interactions only between nearest neighbor chiral
edge modes. As explained in the text, $\Delta$ increases with $d$, but only
slowly, and is smaller than $2$ for arbitrarily large $d$.}
\label{fig:five}
\end{figure}

If the weight function $W(q_{y}) \le 1$ in Eq.~\ref{scale-dim}, then $\Delta_{m}
< 2$ and backscattering is relevant. To understand the dependence of $W(q_{y})$
on filling factor it is useful to consider $q_{y} a= 0,\pi$, so that $D_{+-}$ is
real and the expression for $W$ simplifies. For $\bbox{q}=0$, smectic elasticity
implies $D_{++} + D_{+-} =0$, so that $W(q_{y}=0) = 0$. When $q_{y}a=\pi$, one
has 
\begin{equation}
D_{+-}(q_{y}a =\pi)  = \sum_{n} \frac{(-1)^{n} a}{ 4 \pi^{2} \ell^{2}} 
[\Gamma(an + a(1-\nu^{*})) - \Gamma(an+a\nu^{*})] .
\end{equation}
Note that $D_{+-}(q_{y}a =\pi)$ vanishes, and the weight function equals $1$ for
$\nu^{*} =1/2$. Provided $\Gamma(y)$ is monotonically decreasing with positive
curvature for $y \gtrsim a$, $D_{+-}(q_{y} \pi) $ will be negative for all
$\nu^{*} < 1/2$, implying $W(q_{y}a=\pi) < 1$. If the weight function is
monotonic in $q_{y}a$, the backscattering interactions will thus be relevant.
Using Eq.~(\ref{elastic}) and Eq.~(\ref{gamma}) we have computed $W(q_{y})$ for
a range of values of $N$, $\nu^{*}$, and $d$, and have always found that it is
indeed monotonic; the typical behavior is illustrated in Fig.~(\ref{fig:three}).
For the sake of definiteness we have ignored the finite width of the ground
subband wavefunction in these calculations. Numerically calculated scaling
dimensions for $m=1$, $N=2$, and $d = 10 \ell$, are plotted in
Fig.~(\ref{fig:four}).
For $\nu^{*} > 1/2$, $W(q_{y}a=\pi) > 1$ so that the electron backscattering
amplitude scaling dimension increases, eventually crossing above $2$, as seen in
Fiq.~(\ref{fig:four}). The dependence of the calculated scaling dimension on the
distance to the screening plane is illustrated in Fig.~(\ref{fig:five}) 
for the case $N=2$ and $\nu^{*}=0.5$. As the distance to the screening plane
increases the weighting function approaches $1$ more rapidly as $q_{y} a $ goes
from $0$ to $\pi$. However the values at $q_{y} a = 0,\pi$ are fixed at $0$ and
$1$ respectively, and the curves are monotonic at all values of $d$. As a result
the scaling dimension is only weakly dependent on $d$ and the interaction
remains relevant for any finite value of $d$. 

The most significant conclusion which follows from this calculation is that for
{\em all} $\nu^{*}$, backscattering across either electron or hole stripes is
relevant, and will destabilize the smectic ground state. The ultimate fate of
the ground state will presumably depend on the relative magnitudes of the
various backscattering interactions. For the interactions considered above the
bare coupling constants will fall rapidly with increasing stripe separation $m$:
\begin{equation}
u_{m} \sim U(a\nu/\ell^{2},ma/\ell^{2})   ,
\end{equation}  
so that $m=1$ will dominate. If each electron stripe is viewed as a 1d
conductor, this is a $2k_{F}$ backscattering interaction, which tends to
drive\cite{fradkin} charge ordering {\em along} the stripe, with wavelength
corresponding to the 1d electron spacing. We thus strongly suspect that for
$\nu^{*} < 1/2$ the smectic will be unstable to the formation of an electron
Wigner crystal, with one electron per unit cell. For large Landau index $N$, the
crystal would be highly anisotropic, compressed along the $x$-direction, with an
aspect ratio proportional to $N$. For $\nu^{*} > 1/2$, though, backscattering
across the hole stripes will dominate, leading to an anisotropic hole Wigner
crystal, with one hole per unit cell. In either crystal phase there will, in
contrast to the smectic case, be an energy gap, $E_{g}$, for single particle
excitations. Provided the crystalline order is pinned by the boundaries, these
Wigner crystal phases should have vanishing dissipative conductivities
$\sigma_{xx}$ and $\sigma_{yy}$. However, the hole Wigner crystal will have an
extra Landau level edge state. The quantized Hall conductances of electron and
hole Wigner crystal states will be, $\sigma_{xy} = \lbrack\nu\rbrack e^{2}/h$
and $ \lbrack\nu + 1\rbrack e^{2}/h$ respectively.

\begin{figure}
\begin{center}
\epsfxsize=8.5cm
\epsffile{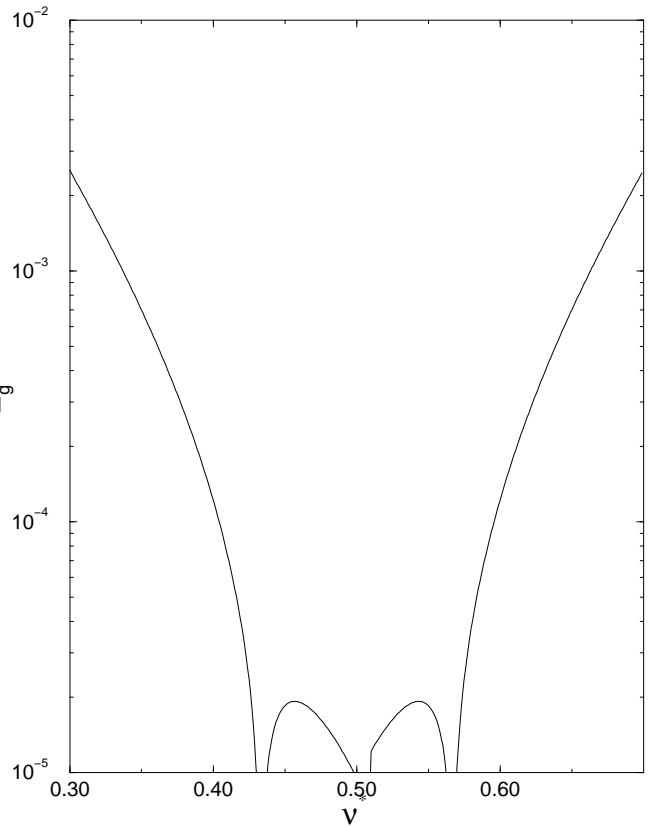}
\end{center}
\caption[]{Estimated single-particle energy gap of the anisotropic Wigner
crystal state, $E_{g}$, as a function of partial filling factor $\nu^{*}$ for a
model with orbital Landau level index $N=2$ and distance to screening plane
$d=10 \ell$. For this model the bare backscattering matrix element vanishes for
$\nu^{*} \sim 0.43$ and $\nu^{*} \sim 0.57$. These results were obtained with
the choice $E_{c} = 0.3 e^{2}/\ell$. The energy gaps are in units of
$e^{2}/\epsilon \ell$ which has a typical value $\sim 100 {\rm K}/k_{B}$.
$E_{g}/k_{B}$ is smaller than $\sim 10 {\rm mK}$, the base temperature scale for
a dilution fridge, for $0.4 \lesssim \nu^{*} \lesssim 0.6$. $E_{g}/k_{B}$
approaches $\sim 1 {\rm K}$, the energy gap observed on reentrant integer
quantum Hall plateaus, for $\nu^{*} \sim 0.25$ and $\nu^{*} \sim 0.75$.}
\label{fig:six}
\end{figure}

Of considerable interest is the {\em magnitude} of the Wigner crystal gap as a
function of $\nu^{*}$. With knowledge of the dimensionless backscattering
interaction, $u$, this gap can be estimated by integrating the RG flow
equations. Specifically, under an RG transformation, the energy gap should
rescale as,
\begin{equation}
E_{g}(u) = b^{-1} E_{g}(b^{2 - \Delta} u )   ,
\end{equation}
with $\Delta = \Delta_{1}$. When the interaction becomes of order one,
$b^{2-\Delta} u =1$, the energy gap should be roughly equal to the
characteristic Coulomb energy, $E_{c}$, giving,
\begin{equation}
E_{g}(u) = (U/E_{c})^{1/(2-\Delta)} E_{c}  ,
\end{equation}
with $U = u E_{c}$ the (dimensionful) backscattering strength. The $\nu^{*}$
dependence of the gap enters both through $U$, which is extremely small for
$\nu^{*}$ near $1/2$ because of the long period of the stripe lattice, and the
scaling dimension, $\Delta$, which is maximal at $\nu^{*} = 1/2$. [For
$\nu^{*}>1/2$ the same applies to backscattering across hole stripes.] Both
effects conspire to {\em strongly} reduce the gap magnitude near half-filling.
Using the above estimates, it is possible to obtain the $\nu^{*}$ dependence of
the gap explicitly. Taking $E_{c} = 0.3 e^{2}/\ell$, the order of the maximum
correlation energy per electron in a partially filled Landau level, the
resulting gap for $N=2$ and $d=10 \ell$ is shown in Fig.~(\ref{fig:six}). 
Notice that the Wigner crystal gap plummets rapidly to extremely small values
near $\nu^{*} =1/2$, dropping below the range accessible to dilution fridges
over the filling factor range $0.4 \lesssim \nu^{*} \lesssim 0.6$, where
anisotropic transport is observed in low-temperature experiments. In this region
the Wigner crystal states will be inaccessible (melted at experimental
temperatures), and the anisotropic transport of the smectic phase should be
unmasked. Outside of this range, the Wigner crystal will be pinned by even weak
impurities, resulting in quantized Hall plateaus. For $\nu^{*} =0.3$, the gap
values estimated here are typical\cite{lilly,du} of those found on the reentrant
integer quantum Hall plateaus which bracket the anisotropic transport regimes.

We remark that electron and hole Wigner crystal states are also the ground
states in the Hartree-Fock approximation.\cite{jungwirth,fradkin,macdunpub} In
that approximation, however, the gaps are orders of magnitude larger $\sim E_{c}
\simeq 0.3~e^{2}/\epsilon\ell$ over a wide-range of filling factors. The
Hartree-Fock approximation is expected to be reasonably accurate for the nearly
classical Wigner crystal states which occur in the tails of $N \leq 1$ Landau
levels. Evidently quantum fluctuations have a larger importance for these $N
\geq 2$ crystal states.

\section{Anisotropic Transport Properties}

Transport near $\nu^{*}=1/2$ in the smectic regime will be strongly influenced
by impurities, which are in fact necessary to get {\em any} transport in the
``hard" $y$-direction. The dominant effect will presumably come from impurity
scattering across electron or hole stripes, with the latter being the bottleneck
when $\nu^{*} < 1/2$ and the former when $\nu^{*}>1/2$. For weak impurity
scattering it is possible to examine their effects perturbatively. Consider for
example impurity scattering across electron stripes,
\begin{equation}
H_{imp} = \int_{x} \sum_{n} \xi_{n}(x) e^{i(\phi_{n+} - \phi_{n-})} + h.c.  ,
\end{equation}
with $\xi(x)$ a complex random potential. Taking $\xi_{n}(x)$ to be uncorrelated
and Gaussian, 
\begin{equation}
[\xi_{n}^{*}(x) \xi_{n^{\prime}}(x^{\prime})]_{ens} = {\cal D}
\delta_{nn^{\prime}} \delta(x-x^{\prime}) ,
\end{equation}
a simple RG perturbative in the variance ${\cal D}$ is possible. One finds,
$\partial {\cal D}/\partial t = (3 - 2 \Delta_{e} ) {\cal D}$, with the scaling
dimension of the operator $e^{i(\phi_{+} - \phi_{-})}$ given by,
\begin{equation}
\Delta_{e} = \int_{-\pi}^{\pi} { {d(qa) } \over {2\pi} } W(q_{x}=0,q)   .
\end{equation}
Here $W$ is the {\em same} ``weight" function as in Eq.~\ref{scale-dim}. The
filling factor dependence of $\Delta_{e}$ can be understood from considerations
similar to those for the backscattering amplitudes detailed in the previous
section. For 1D non-interacting electrons $\Delta_{e} =1$, so that disorder is
relevant and eventually leads to localization. For the smectic we can estimate
$\Delta_{e}$ as a function of filling $\nu^{*}$; the result of this calculation
was included in Fig.~(\ref{fig:four}). At all $\nu^{*}$ impurity scattering
across either electron or hole stripes is more relevant than in the
non-interacting electron case.

In the strict zero temperature limit, we thus expect that impurities (aided by
interactions) will ultimately drive localization for all $\nu^{*}$, except right
{\em at} the $\nu^{*} =1/2$ plateau transition. However, samples in which
quantum-Hall smectic physics is observed have extremely weak impurity
scattering, so that it might be possible to ignore localization effects at
accessible temperatures. More specifically, consider the dimensionless disorder
strength, $D = {\cal D} \ell/E_{c}^{2}$, with $E_{c}$ the Coulomb energy scale.
Provided $D <<1$, there should be a large temperature range over which impurity
backscattering can be treated perturbatively and localization effects ignored.
To see this, it is convenient to introduce an {\em effective}
temperature-dependent disorder strength that follows from the RG; $D_{\rm
eff}(T) = (T/E_{c})^{2\Delta_{e} -3} D$, which increases upon cooling. Provided
$D_{\rm eff}(T) < 1$, localization effects should be negligible, and Boltzmann
transport should be operative. 

A key parameter in a Boltzmann approach is the impurity scattering rate
$\Gamma_{e}$ ($\Gamma_{h}$) across an electron (hole) stripe. Within a simple
free-fermion Golden-rule calculation one expects $\Gamma_{e}^0 = cD E_{c}$ (with
$c$ an order one constant), which is independent of temperature. But under the
RG transformation the scattering rate rescales as,
\begin{equation}
\Gamma_{e} (D,T) = b^{-1} \Gamma_{e}(b^{3-2\Delta_{e}} D, bT)   .
\end{equation}
Running the RG until $bT = E_{c}$ gives, $\Gamma(D,T) = (T/E_{c}) \Gamma(D_{\rm
eff},E_{c})$. Using the free-fermion result at $T=E_{c}$ one has
\begin{equation}
\Gamma_{e}(T) = cT D_{\rm eff}(T) = \Gamma_{e}^0 \,
(T/E_{c})^{2\Delta_{e} -2}  .
\label{Rate-T}
\end{equation}
This should be valid provided that $D_{\rm eff} < 1$. For a non-interacting 1D
electron gas $\Delta_{e} =1$ so that $\Gamma_{e}$ is temperature independent. In
contrast, Luttinger liquid effects in the quantum Hall smectic give a temperature
dependence to the Boltzmann scattering rate - generally increasing upon cooling.
Equivalently, the impurity mean free path varies with temperature, in marked
contrast to low temperature metallic transport.

In the Boltzmann approach to transport in the quantum Hall smectic that we
develop below, quantum interference effects between successive inter-edge
impurity backscattering events are ignored. This is valid provided, $\Gamma_{e}$
is not large compared to $\Gamma_{\phi}$, where $\Gamma_{\phi}$ is the electron
phase breaking rate. Within a single chiral edge mode, forward scattering
interactions will rapidly dephase an electron. A simple perturbative calculation
for the electron self energy is expected to give the form; $\Gamma_{\phi} =
c^{\prime} T u_{f}^{2}$ with $u_{f}$ a dimensionless forward scattering
amplitude and $c^{\prime}$ of order one. Since $u_{f}$ is also of order one this
implies, $\Gamma_{\phi} = c_{\phi} T$. Comparing with Eq.~\ref{Rate-T}, one sees
that it is thus legitimate to ignore interference between successive impurity
backscattering events provided $D_{\rm eff}(T)$ is not large compared to one.
For temperatures low enough that $D_{\rm eff}(T)$ is large, quantum interference
effects cannot be neglected and one expects an onset of localization (except
right {\em at} $\nu^{*} = 1/2$). In strong field, the leading one-loop weak
localization effects will not be operative, so that two-loop interference
processes will drive the localization.

With this preamble in hand, we proceed to develop a semiclassical Boltzmann
transport theory for the quantum Hall smectic phase. We assume that the charge
density wave itself is pinned and immobilized by both the edges of the sample
and weak impurities which couple to the electrons within the stripes. In this
case, collective sliding motion of the charge-density will be absent, and the
electrical transport will be dominated by single-particle inter-edge electron
tunneling.  It is convenient to characterize the non-equilibrium
current-carrying state by separate local steady state chemical potentials,
$\mu_{n\pm}$, for left and right going electrons in each stripe. Due to the
discrete translational symmetry of the smectic, the steady state chemical
potential must increase by $e E_{y} a$ upon translation by one period, with
$E_{y}$ the $y$-component of a (uniform) electric field. Taking the zero of
chemical potential as the center of the $n=0$ electron stripe, we can thus
write,
\begin{eqnarray} 
\mu_{n,+} &=&  n e E_{y} a + \mu/2 \nonumber ,\\
\mu_{n,-} &=&  n e E_{y} a - \mu/2 .
\label{chempot}
\end{eqnarray}
Here the chemical potential drops across electron and hole stripes are $\mu$ and
$e E_{y} a - \mu$ respectively. An electric field $E_{x}$ in the $\hat x$
direction, induces a steady flow in momentum space which moves each electron
stripe to smaller $y$ (for $E_{x} >0$), lowering the chemical potential on
right-going edges and raising it on left-going edges. This disequilibrium
induces a tunneling current across both electron and hole stripes, which
attempts to restore equilibrium:
\begin{eqnarray}
\dot{\mu}_{n,+} &=& - e E_{x} v_{F} + \frac{\mu_{n,-} - \mu_{n,+}}{\tau_{e}} +
\frac{\mu_{n+1,-} - \mu_{n,+}}{\tau_{h}} , \nonumber \\
\dot{\mu}_{n,-} &=& e E_{x} v_{F} + \frac{\mu_{n,+} - \mu_{n,-}}{\tau_{e}} +
\frac{ \mu_{n-1,+} - \mu_{n,-}}{\tau_{h}}.
\label{tdbe}
\end{eqnarray}
Here, we have introduced inter-edge scattering times, related to the rates above
via: $\Gamma_{e} = 1/\tau_{e}$ and $\Gamma_{h} = 1 / \tau_{h}$, for tunneling
across electron and hole stripes, respectively. The electric field, $E_{x}$,
induces a drift in the wavevector of the electrons in each chiral edge mode,
$\hbar \dot{k} = -eE_{x}$. In Eq.~(\ref{tdbe}) $v_{F}$ is a ``Fermi velocity",
which relates changes in the edge chemical potential to wavevector: $v_{F} =
\partial \mu / \partial k$. This velocity is determined by the ``onsite" piece
of the smectic elastic constants as, $v_{F} = 2\pi \ell^{2} D_{++}(q_{x} =0,
n=0) $. 

In the steady state $\dot{\mu}_{n,\pm} = 0$ so that 
\begin{equation}
\mu (\tau_{e}^{-1} + \tau_{h}^{-1}) = \frac{e E_{y} a}{\tau_{h}} - e E_{x} v_{F},
\label{mu}
\end{equation}
relating the unknown parameter $\mu$ to the electric fields. The current in the
$\hat x$ direction is due to the imbalance between left-going and right-going
electrons in each stripe 
\begin{equation}
I_{x} = \frac{e^{2}}{h} \left[\frac{L_{y}}{a}\right] (-\mu/e).
\label{ix}
\end{equation} 
In Eq.~(\ref{ix}) the contribution from each stripe is given by the familiar
expression for the quantum Hall current and the factor in square brackets is the
number of electron stripes in a sample with width $L_{y}$. The current in the
$\hat y$ direction is equal to the tunneling current across the hole stripes: 
\begin{equation}
I_{y} = \frac{e L_{x}}{v_{F} h} \frac{eE_{y} a - \mu}{\tau_{h}}.
\label{iy}
\end{equation}
with $L_{x}$ the sample width. The first factor on the right-hand-side of
Eq.~(\ref{iy}) is the charge per unit energy in a chiral 1D electron system of
length $L_{x}$.

Inserting Eq.~(\ref{mu}) in Eq.~(\ref{ix}) and Eq.~(\ref{iy}) to eliminate $\mu$
gives the desired expressions for the conductivity matrix,
\begin{eqnarray}
\sigma_{xx} &=& \frac{e^{2}}{h} \frac{v_{F} \tau_{e} \tau_{h}}{a (\tau_{e} +
\tau_{h})}, \nonumber \\
\sigma_{yy} &=& \frac{e^{2}}{h} \frac{a}{v_{F}(\tau_{e} + \tau_{h})}, \nonumber\\
\sigma_{yx} &=& - \sigma_{xy} = \frac{e^{2}}{h} \left(\lbrack\nu\rbrack +
\frac{\tau_{e}}{\tau_{e} + \tau_{h}}\right).
\label{sigma}
\end{eqnarray}
Inverting the conductivity matrix gives the following expressions for the
resistivities:
\begin{eqnarray}
\rho_{\rm easy} &=& \frac{h}{e^{2}} \, \frac{1}{\tau_{e} \left(\lbrack\nu\rbrack
+1\right)^{2} + \tau_{h} \lbrack\nu\rbrack^{2}} \, \frac{a}{v_{F}}\nonumber\\
\rho_{\rm hard} &=& \frac{h}{e^{2}} \, \frac{1}{\tau_{e} \left(\lbrack\nu\rbrack
+1\right)^{2} + \tau_{h} \lbrack\nu\rbrack^{2}} \, \frac{v_{F} \tau_{e}
\tau_{h}}{a}\nonumber\\
\rho_{\rm hall} &=& \frac{h}{e^{2}} \frac{1}{\tau_{e} \left(\lbrack\nu\rbrack
+1\right)^{2} + \tau_{h} \lbrack\nu\rbrack^{2}} \, \left(\lbrack\nu\rbrack
+1\right) \tau_{e} + \lbrack\nu\rbrack \tau_{h},
\label{rho}
\end{eqnarray}
where $\rho_{\rm easy}=\rho_{xx}$, $\rho_{\rm hard}=\rho_{yy}$, and $\rho_{\rm
hall}=\rho_{xy}$. 

Eq.~(\ref{rho}) relates the dissipative and Hall resistivities to the two
scattering rates, $\Gamma_{e}$ and $\Gamma_{h}$. The dependencies on temperature
and filling factor $\nu^{*}$ enter through these scattering rates, in the form
established above:
\begin{eqnarray}
\Gamma_{e} &=& \frac{1}{\tau_{e}} = \Gamma_{e}^{(0)} (k_{B} T/ E_{c})^{2
\Delta_{e}-2} , \nonumber \\ 
\Gamma_{h} &=& \frac{1}{\tau_{h}} = \Gamma_{h}^{(0)} (k_{B} T /E_{c})^{2
\Delta_{h}-2}.
\label{gammadis}
\end{eqnarray} 
Here, the free-fermion scattering rates across electron and hole stripes,
$\Gamma_{e}^{(0)}$ and $\Gamma_{h}^{(0)}$, depend on the impurity scattering
strength (and $\nu^{*}$) but {\em not} the temperature. $E_{c}$ is the Coulomb
energy scale which serves as a high energy cutoff. The scaling dimensions,
$\Delta_{e}$ and $\Delta_{h}$, depend sensitively on $\nu^{*}$ as shown in
Fig.~(\ref{fig:four}). As we shall see, these equations describe much of the
phenomenology\cite{lilly,du,shayegan} of transport in quantum Hall stripe
states. 

Remarkably, for $\nu^{*} =1/2$ this theory makes two parameter free quantitative
predictions:
\begin{equation}
\rho_{\rm easy} \, \rho_{\rm hard} = (h/e^{2})^{2}
\frac{1}{\left[\left(\lbrack\nu\rbrack +1\right)^{2} +
\lbrack\nu\rbrack^{2}\right]^{2}} , 
\label{easyhardhalf}
\end{equation}
and 
\begin{equation}
\rho_{\rm hall} = \frac{h}{e^{2}} \frac{2\lbrack\nu\rbrack
+1}{\left(\lbrack\nu\rbrack +1\right)^{2} + \lbrack\nu\rbrack^{2}}.
\label{hallhalf}
\end{equation}
Notice that the scattering times have completely dropped out of these
expressions! Interestingly, the Hall resistivity at $\nu = \lbrack\nu\rbrack
+1/2 $ is predicted to deviate noticeably from (the classical value) $
(h/e^{2})/(\lbrack\nu\rbrack +1/2)$. The most extensive experimental data is
for $\lbrack\nu\rbrack =4$. In this case, the value predicted for the product
of $\rho_{\rm easy}$ and $\rho_{\rm hard}$ appears to agree with the published
data to within better than a factor of two, provided one accounts for the
particular current paths\cite{simon} appropriate for the sample geometry.
Experimental verification of the predicted $\lbrack\nu\rbrack$ dependence of
this product would help establish the efficacy of this transport theory.

At $\nu^{*} = 1/2$, Eq.~(\ref{rho}) predicts a weak temperature dependence of
the dissipative resistivities. Specifically, due to Luttinger liquid effects
which drive an {\em enhancement} of the inter-edge scattering rate upon cooling
(since $\Delta_{e/h} < 1$ at $\nu^{*} =1/2$), the resistivity in the hard
direction should drop slowly with cooling whereas $\rho_{\rm easy}$ should rise.

It is interesting to consider the predicted dependence of the resistivities on
filling factor. For $\nu^{*}< 1/2$, the electron stripes are narrower than the
hole stripes and a free-fermion evaluation of the relaxation times would give
$\tau_{h} > \tau_{e}$. Since $\Delta_{e}$ decreases and $\Delta_{h}$ increases
with increasing $1/2-\nu^{*}$, the relaxation rate ratio is expected to increase
beyond its free-fermion value at lower temperatures. For $\tau_{h} \gg \tau_{e}$
we have that $\rho_{\rm hard} = (h/e^{2}) (v_{F} \tau_{e} /a)
/{\lbrack\nu\rbrack}^{2}$. Since $\tau_{e}$ decreases ever more rapidly upon
cooling for larger $1/2 - \nu^{*}$, the hard resistivity is expected to be large
at experimental temperatures only over a narrow interval surrounding $\nu^{*}
=1/2$. Backscattering interactions ignored in this Boltzmann transport theory
will only tend to enhance this effect, acting in concert with impurity
scattering. 

In the same regime of filling factor, with $\tau_{h} >> \tau_{e}$, the Hall
resistivity approaches $(h/e^{2})/\lbrack\nu\rbrack$. Moreover, one has
$\rho_{\rm easy} = (h/e^{2})/{\lbrack\nu\rbrack}^{2} (v_{F} \tau_{h}/a)$ in
this limit. Thus, $\rho_{\rm easy}$ also decreases with $1/2-\nu^{*}$, because
of both bare matrix element and scaling dimension tendencies. Interestingly, for
$\nu^{*} \lesssim 0.4$ the scaling dimension for scattering across hole stripes,
$\Delta_{h}$, becomes {\em larger} than one (see Fig.~(\ref{fig:four}). This
implies that $\Gamma_{h}$ actually decreases upon cooling in this regime,
strongly enhancing the one-dimensional nature of the electron stripes and
driving localization. But for $\nu^{*} \lesssim 0.4$ one really must include the
strong effects of electron (backscattering) interactions which drive the Wigner
crystal instability - again, this acts in concert with impurity affects. Upon
cooling within this Wigner crystal regime, the dissipative resistivities should
rapidly vanish leaving a quantized Hall resistance. It follows from
particle-hole symmetry that $\tau_{e}(\nu^{*}) = \tau_{h}(1-\nu^{*})$ so similar
conclusions can be reached for transport properties at $\nu^{*} > 1/2$. 

\begin{figure}
\begin{center}
\epsfxsize=8.5cm
\epsffile{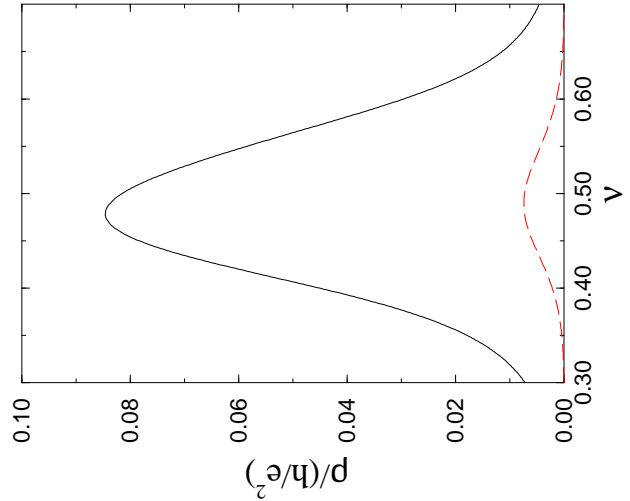}
\end{center}
\caption[]{Linear transport coefficients calculated from the Boltzmann transport
theory for $N=2$ and $d = 10 \ell$, using a model of $\delta$-correlated
disorder. The disorder strength was chosen to give a ration of hard to easy
direction resistivities equal to $10$ at $\nu^{*}=0.5$ and the temperature was
chosen to be $100$ times smaller than the microscopic interaction strength as
explained in the text. Interaction and localization effects neglected in this
Boltzmann theory are expected to strongly suppress these resistivities at low
temperatures outside of the interval $0.4 \lesssim \nu^{*} \lesssim 0.6$.}
\label{fig:seven}
\end{figure}

In Fig.~(\ref{fig:seven}),
we plot the $\rho_{\rm easy}$ and $\rho_{\rm hard}$ filling factor dependencies
predicted by this model for $\lbrack\nu\rbrack = 4$ and $\delta$-correlated
disorder model. Here we have taken $T=0.003 e^{2}/\ell$ and $E_{c} = 0.3 e^{2}/
\ell$. The disorder strength was chosen to give $\rho_{\rm hard}/\rho_{\rm
easy}=10$ at $\nu^{*} = 0.5$. Notice that $\rho_{\rm easy}$ has a maximum at
$\nu^{*}=1/2$, not the minimum seen in many experiments. Within the Boltzmann
theory this feature depends on the details of the disorder model used; models
with only small angle scattering at zero magnetic field tend to give electron
relaxation times which decrease and hole relaxation times which increase more
rapidly with $1/2 - \nu^{*}$, changing the shape of these curves. In addition,
current-path corrections\cite{simon} might be essential in producing the
apparent $\rho_{\rm easy}$ minimum at $\nu^{*}=1/2$ in experiments. No plausible
disorder model in this theory gives $\rho_{\rm hard}$ results which drop to zero
as strongly with increasing $1/2 - \nu^{*}$ as in experiments; we believe that
backscattering interactions and localization, both effects neglected here, are
playing an important role in driving the resistivities to small values away from
$\nu^{*} =1/2$.

The transport theory described above can be readily generalized to account for
the non-linear transport features, which are present experimentally - notably at
$\nu^{*}=1/2$. The voltage drop across a stripe, given by $V_{y} = aE_{y}/2$ at
$\nu^{*} = 1/2$, can be readily incorporated into the RG scaling approach for
the scattering rates. Not surprisingly, the resulting dependence on voltage is
the same as that on temperature obtained above:
\begin{eqnarray}
\Gamma_{e} &\sim& \Gamma_{e}^{(0)} (V_{y}/ E_{c})^{2 \Delta_{e}-2} \nonumber\\
\Gamma_{h} &\sim& \Gamma_{h}^{(0)} (V_{y} /E_{c})^{2 \Delta_{h}-2} .
\label{nonlintrans}
\end{eqnarray} 
This expression is valid in low temperature or high voltage limits, with $k_{B}T
\ll V_{y}$. The non-linear differential resistivity in the hard direction can
now be obtained by using the expression for the tunneling current across stripes
from Eq.~(\ref{rho}): $I_{y} = (e^{2}/h) \Gamma_{e} V_{y} L_{x}/v_{F}$. One
thereby obtains,
\begin{equation} 
\frac{\partial V_{y}}{\partial I_{y}} \sim I_{y}^\alpha  ,
\label{diffresisty}
\end{equation}
with an exponent $\alpha = 2(1-\Delta_{e})/(2\Delta_{e} -1)$. Using the value of
$\Delta_{e} = 0.756$ calculated from theory at $\nu^{*} =1/2$ for the case
$[\nu]=4$ and $d =10 \ell$ (see Fig.~[\ref{fig:four}], gives the estimate
$\alpha = 0.93$. Calculations for models with more remote screening planes will
give larger values for $\Delta_{e}$ (but always smaller than one as explained
above) and smaller positive values for $\alpha$. Notice that a positive exponent
implies an enhancement of the hard axis resistivity when driven non-linear -
consistent with the experimental findings in Ref. 1. This increase in
resistivity is due to a voltage suppression of the correlation induced
interlayer tunneling enhancement - and as such is a rather direct experimental
indication of non-trivial Luttinger liquid correlations of the chiral edge
channels. 

A similar calculation can be performed for non-linearities in the easy axis
current. At $\nu^{*} = 1/2$ \, $V_{x} \propto \Gamma_{e} I_{x}$ where $I_{x}$ is
the easy-direction current. In this case $V_{y} \propto \nu I_{x}$ is the Hall
voltage. It follows that 
\begin{equation} 
\frac{\partial V_{x}}{\partial I_{x}} \sim I_{x}^\beta  
\label{diffresistx}
\end{equation}
with an exponent $\beta = 2(\Delta_{e} -1)$ For $N=2$ and $d=10 \ell$, we obtain
$\beta = -0.48$; smaller negative values are found for models with more remote
screening planes. In this theory the easy-direction resistivity is suppressed
when driven non-linear with current, a property which is also
consistent\cite{lilly} with experimental findings. We emphasize that these
powerlaws hold only in the low-temperature or high-voltage limits; a careful
comparison of the theoretical dependence on voltage to temperature ratio with
experiment could provide a strong test of this transport theory.

\section{Summary}

Recent experiments\cite{lilly,du,shayegan} have established a consistent set of
transport properties for high-mobility two-dimensional electron systems with
high orbital index ($ N \ge 2 $) partially filled Landau levels which differ
from those in the low orbital index ($ N \le 1 $) fractional quantum Hall
regime. At large $N$, the dissipative resistivities are large, strongly
anisotropic, and non-linear for $0.4 \lesssim \nu - \lbrack\nu\rbrack \lesssim
0.6$ within each Landau level. This anisotropic transport regime is bracketed by
regions of reentrant integer quantum Hall plateaus. In this paper we have
presented a theory which is able to account for most features of these
experiments. The theory starts from the unidirectional charge-density-wave
(smectic) state of Hartree-Fock theory\cite{fogler,moessner} in which the
electrons reside in periodically spaced stripes with a spontaneously chosen
orientation. Forward and backscattering interactions, neglected in the
Hartree-Fock theory, are included by retaining the low energy electron
excitations at the stripe edges. These form a set of coupled 1D chiral modes,
easily described with bosonization techniques. We find that: i) for smectic
states in quantum Hall systems, the chiral boson degrees of freedom coincide
with stripe position and width degrees of freedom; ii) backscattering
interactions which drive the system toward electron or hole Wigner crystal
states are always relevant, but only below inaccessibly low temperatures in the
anisotropic transport regime; and iii) a semiclassical Boltzmann transport
theory for the smectic state is able to account for the magnitude of the
anisotropic dissipative resistivities and for the sign of the non-linearities
which appear at higher transport currents.

\section{Acknowledgements}

We would like to acknowledge insightful conversations with Herb Fertig, Michel
Fogler, Tomas Jungwirth, Jim Eisenstein, Eduardo Fradkin, and Steve Girvin and
the stimulating environment provided by the ITP, where this research was
initiated. M.P.A.F. is grateful to the NSF for generous support under grants
DMR-97-04005, DMR95-28578 and PHY94-07194. A.H.M. is grateful for support under
NSF grant DMR-97-14055.


\begin{references}
\bibitem{Finite-T} At non-zero temperatures an unbinding of disclinations will
destroy the translational LRO, but algebraic rotational order can survive.
\bibitem{remarkotherterm} At $\nu^{*}=1/2$ there is an additional class of
momentum-conserving low-energy processes in which one particle scatters across
an electron stripe and the second particle across a hole stripe. Calculations
similar to those discussed here show that these are irrelevant in a
renormalization group sense.
\end{references}
\end{document}